\documentclass[11pt,fleqn]{article}

\usepackage{a4}
\usepackage{amstext}
\usepackage{amsfonts}
\usepackage{amssymb}
\usepackage{color}
\usepackage{epsfig}

\begin{document}


\begin{flushright}
HU-EP-09/10 \\
SFB/CPP-09-24
\end{flushright}

\begin{center}

{\huge \bf Cautionary remarks on the moduli space}

{\huge \bf metric for multi-dyon simulations}

\vspace{0.5cm}

\textbf{Falk Bruckmann} \\
Institut f\"ur Theoretische Physik, Universit\"at Regensburg, D-93040 Regensburg, Germany \\
Institut f\"ur Theoretische Physik III, Universit\"at Erlangen, D-91058 Erlangen, Germany\\
\texttt{falk.bruckmann@physik.uni-regensburg.de}

\vspace{0.5cm}

\textbf{Simon Dinter, Ernst-Michael Ilgenfritz, Michael M\"uller-Preussker, Marc Wagner}
Humboldt-Universit\"at zu Berlin, Institut f\"ur Physik, Newtonstra{\ss}e 15, D-12489 Berlin, Germany  \\
\texttt{dinter@physik.hu-berlin.de} \\
\texttt{ilgenfri@physik.hu-berlin.de} \\
\texttt{mmp@physik.hu-berlin.de} \\
\texttt{mcwagner@physik.hu-berlin.de}

\vspace{0.5cm}

March 17, 2009

\end{center}

\vspace{0.1cm}

\begin{tabular*}{16cm}{l@{\extracolsep{\fill}}r} \hline \end{tabular*}

\vspace{-0.4cm}
\begin{center} \textbf{Abstract} \end{center}
\vspace{-0.4cm}

We perform a detailed numerical investigation of the approximate moduli 
space metric proposed by Diakonov and Petrov \cite{Diakonov:2007nv}
for a confining model of dyons. 
Our findings strongly indicate that only for a small number of dyons at 
sufficiently low density this metric is positive definite 
-- and, therefore, a valid moduli space metric -- 
throughout a considerable part of configuration space.
This poses strong limitations on results obtained by an unrestricted integration over collective coordinates in this model. It also indicates that strong correlations between collective coordinates will be essential for the physical content of a dyon model, which could be exhibited by a suitable simulation algorithm.

\begin{tabular*}{16cm}{l@{\extracolsep{\fill}}r} \hline \end{tabular*}

\thispagestyle{empty}


\newpage

\setcounter{page}{1}

\section{\label{sec:INTRODUCTION}Introduction}

The semiclassical approximation of path integrals around non-trivial saddle 
points (originating from the work of Callan, Dashen and
Gross~\cite{Callan:1977gz,Callan:1978bm} and later adapted to finite temperature
by Gross, Pisarski and Yaffe~\cite{Gross:1980br})
is a prominent, but not undisputed~\cite{Witten:1978bc,Horvath:2002yn,Ahmad:2005dr} 
semianalytic approach to calculate non-perturbative effects in quantum field 
theories. In gauge theories at zero temperature the corresponding semiclassical 
topological objects used to be instantons~\cite{Belavin:1975fg}, four-dimensional 
selfdual lumps of action density carrying one unit of topological charge.
Their use in a semiclassical context was opened by the seminal paper by 
't Hooft~\cite{'tHooft:1976fv}, where the fluctuation determinant in the background 
of such objects was computed. However, the resulting distribution with respect to 
the size parameter $\rho$ is obviously unphysical for large instantons. 
Empirical interactions~\cite{Ilgenfritz:1980vj,Munster:1981zn,Diakonov:1983hh} 
have been added to the first instanton gas model~\cite{Callan:1977gz,Callan:1978bm}
eventually leading to the instanton liquid model~\cite{Shuryak:1981ff},
where the average size is fixed to $\bar{\rho} = 1/3 \, \textrm{fm}$
(along with a density of $n^{-1/4} = 1 \, \textrm{fm}$) to roughly match 
phenomenological requirements.

In pure gauge theories at finite temperature the Polyakov loop is the order 
parameter for confinement. That there is a close relation between the Polyakov 
loop average and the dominating classical solutions has been fully realized with 
the discovery of calorons with non-trivial holonomy 
by Kraan and van~Baal~\cite{Kraan:1998pm} and Lee and Lu~\cite{Lee:1998bb}.
From this perspective, the previously known caloron solutions due to Harrington 
and Shepard~\cite{Harrington:1978ve} now appear as the limiting case of trivial 
holonomy\footnote{For a dyon model based on these solutions 
see~\cite{Simonov:1995cr}.}.
The eigenvalues of the untraced asymptotic Polyakov loop -- called holonomy --
are responsible for the dissociation of calorons into constituents, $N$ for 
gauge group SU($N$). These are BPS monopoles with electric charges equal or 
opposite to their magnetic charges. We will refer to them as dyons (for reviews 
cf.~\cite{Bruckmann:2007ru,Diakonov:2008sg}).

Hence, a semiclassical model of finite temperature gauge theory should be based 
on dyons. 
If the holonomy is related to the order parameter, all types of dyons are 
of equal mass in the confined phase but expected to split into light and heavy
ones in the deconfined phase. Dyon gauge fields combined to intermediate size 
calorons do not generate confinement at trivial holonomy, whereas at maximally 
non-trivial holonomy the corresponding static potential is believed to be 
linearly rising. This has been demonstrated numerically in~\cite{Gerhold:2006sk}. 
As size distribution in the confinement phase a generalization of the instanton 
distribution was adopted that was inspired by the evaluation of the fluctuation 
determinant in refs.~\cite{Diakonov:2004jn,Zarembo:1995am}.
The suppression of heavy dyons due to the non-vanishing Polyakov loop above the 
critical temperature could be the mechanism behind the decrease of the topological 
susceptibility and the vanishing of the chiral condensate~\cite{Bruckmann:2009ne}. 
In contrast to this, the chiral condensate under (unphysical) periodic boundary 
conditions is non-vanishing, an effect seen in various studies~\cite{Gattringer:2002tg,Bornyakov:2008bg,Bornyakov:2008im,Stephanov:1996he,Bilgici:2008qy} that is likely to reflect the 
presence of light dyons.

Here we present results of a preparatory study of SU(2) gauge theory
trying to obtain numerical insight into a dyon model recently developed by 
Diakonov and Petrov~\cite{Diakonov:2007nv,Diakonov:2008sg,Diakonov:2008rx}.
In this work, the role of the moduli space metric was emphasized and incorporated.
The authors have generalized the known form of the metric of same kind dyons and 
of opposite kind dyons to a general metric valid at large distances
in the parameter space of an ensemble of selfdual dyons. 
The problem of including dyons of opposite topological charge was postponed, which means that the model is a rather crude approximation.
The authors of~\cite{Diakonov:2007nv} attempted to take the effects of a mixture 
of selfdual and antiselfdual objects into account by multiplying some of their 
results by factors of $2$ or $\sqrt{2}$ assuming negligible interactions between 
the two systems of selfdual and antiselfdual dyons.

The resulting moduli space metric determinant consists of Coulomb-like terms and 
has been treated analytically by rewriting the corresponding grand-canonical 
ensemble into an equivalent quantum field theory both of bosons and fermions, 
resembling Polyakov's famous work~\cite{Polyakov:1976fu} 
showing Abelian confinement. The relations between physical quantities such as 
the string tension and the critical temperature obtained from this model are rather 
impressive, when comparing them with lattice results. However, 
in order to match phenomenological values, the model needs to be pushed to rather 
high densities (i.e.\ short distances between dyons), which is difficult to 
reconcile with the diluteness assumption of the underlying semiclassical approach. 
It should be mentioned that a similar problem afflicts the instanton liquid model 
as well (when choosing size and density as quoted above).

Our final goal is to complement the analytical treatment of the dyon model 
presented in ref.~\cite{Diakonov:2007nv} by numerical simulations of the proposed 
moduli space metric together with interactions stemming from the gluonic fluctuation
determinant,
the Faddeev-Popov determinant and the action itself. The preparatory studies to be
presented here will demonstrate that the approximate moduli space metric
from~\cite{Diakonov:2007nv} violates the fundamental requirement of being positive 
definite in an overwhelming part of dyon configuration space, when explored at 
densities needed to match phenomenology. 
We conclude that this moduli space metric can only be used in a dyon model with other terms added, or if correlations between the dyons are built in that lead to the positivity of the metric. A suitable simulation algorithm should guarantee that.
This would result in complicated multi-dyon correlations and create a non-trivially related behavior of various observables and correlators.

This paper is organized as follows. In section~\ref{sec:SECTION2} we briefly recall 
the ingredients of the dyon model in the version of Diakonov and Petrov applied
to SU(2) gauge theory, which is the starting point for our work.
Section~\ref{sec:SECTION3} deals with the spectral properties of the proposed 
moduli space metric. In section~\ref{SEC471} we demonstrate that even a random model of dyons, i.e.\ a model without moduli space metric, induces confinement.
Finally we conclude and give a brief outlook.


\section{\label{sec:SECTION2}Dyon ensembles \`a la Diakonov and Petrov}

\subsection{\label{subsec:holonomy}Holonomy as external parameter}  

We consider pure SU(2) gauge theory at finite temperature $T$, which, as usual, 
is implemented with periodic boundary conditions in the imaginary time direction 
with period $\beta = 1 / T$. There are four kinds of dyons, two selfdual and two 
antiselfdual. 
The untraced Polyakov loop at spatial infinity, also called holonomy, is an 
element of the gauge group. It can be diagonalized 
everywhere\footnote{To be more precise, it can be diagonalized everywhere except 
for the loci of the Dirac strings, if the system is not neutral.}
to $\Omega = \exp(2 \pi i \omega \sigma_3)$ with eigenvalues $e^{+2 \pi i \omega}$ 
and $e^{-2 \pi i \omega}$, where $\omega\in[0,1/2]$. 
Similarly to the non-Abelian adjoint Higgs 
system this gives rise to complementary dyons that have actions proportional to 
$2 \omega$ and $1 - 2 \omega$ and that become static when well-separated. They are 
both selfdual or antiselfdual depending on the sign of their topological charge. 

Following Diakonov and Petrov~\cite{Diakonov:2007nv}, throughout this paper we 
focus on the case of maximally non-trivial holonomy, $\omega = 1/4$. This 
implies that all dyons have the same action. The holonomy is then traceless 
and matches the confinement condition $\langle P \rangle = 0$, where 
$P = \textrm{Tr}(\Omega) / 2$.
We adopt the same restriction to purely selfdual systems, considering $K$ dyons of 
the first kind and $K$ dyons of the second kind, i.e.\ the total number of dyons is 
$n_D = 2 K$. These dyons carry equal electric and magnetic charges which can
take the values $\pm 1$ .


\subsection{\label{subsec:metric}The approximate moduli space metric of multi-dyon 
configurations}

In~\cite{Diakonov:2007nv} it has been attempted to construct an approximate 
multi-dyon moduli space metric valid for dyon separations $d \gg 1 / \pi T$. 
The starting point for this construction were the analytically known moduli 
space metric of a single caloron~\cite{Kraan:1998pn}, which is a pair of different 
kind dyons, and a corresponding approximation for pairs of same kind dyons at 
large distances. 
The integration over collective coordinates, which are the dyon positions
$\mathbf{x}_i^m$ \footnote{The phases of the dyons are irrelevant in this context.} 
with $i = 1,\ldots,K$ denoting the dyon index and
$m = 1,2$ denoting first and second kind respectively, is then performed with 
the measure
\begin{eqnarray}
\label{EQN873} \bigg(\prod_{i=1}^K \prod_{m=1}^2 d^3x_i^m\bigg) \sqrt{\textrm{det}(g)} \,  .
\end{eqnarray}
The approximate moduli space metric $g$ is related to a matrix $G$, 
\begin{equation}
\label{EQN001} G_{i,j}^{m,n} \ \  = \ \ \delta^{m,n}\delta_{i,j} \bigg(2\pi + 2\sum_{k=1}^K \frac{1}{d_{i,k}^{m,m+1}} - 2\sum_{k=1, k \neq i}^K \frac{1}{d_{i,k}^{m,m}}\bigg)+ 2\frac{\delta^{m,n}(1-\delta_{i,j})}{d_{i,j}^{m,m}} - 2\frac{\delta^{m,n+1}}{d_{i,j}^{m,m+1}}\,,
\end{equation}
(see appendix~\ref{SEC011}) such that the determinants are related by 
%
\begin{equation}
\label{EQN999} \sqrt{\textrm{det}(g)} \ \ = \ \ \textrm{det}(G) \, .
\end{equation} 
The notation $d_{i,j}^{m,n} = |\textbf{x}_i^m - \textbf{x}_j^n|$ is used for
dyon separations, and dyon type indices $m$ are considered as cyclic, i.e. for 
SU(2) holds $(m=3) \equiv (m=1)$.
We omit the temperature $T$ in most equations, which means that distances are
measured in $1/T$ units. The temperature dependence can easily be restored by 
making this explicit.


\subsection{\label{subsec:missing_positivity}Parameters and their physical values}

Within the model proposed in~\cite{Diakonov:2007nv} the relation between the 
string tension $\sigma$ and the critical temperature $T_c$ of the confinement 
deconfinement phase transition can be derived analytically. It reads for SU(2)
gauge theory:
\begin{eqnarray}
\label{EQN030} & & \hspace{-0.7cm} \sigma \ \ = \ \ \Big(8 \rho T\Big)^{1/2} \quad , \quad T_c \ \ = \ \ \Big(48 \rho T / \pi^2\Big)^{1/4} \, ,
\end{eqnarray}
or equivalently
\begin{eqnarray}
\frac{T_c}{\sqrt{\sigma}} \ \ = \ \ \bigg(\frac{6}{\pi^2}\bigg)^{1/4} \ \ = \ \ 0.883 \, .
\end{eqnarray}
Eqns.\ (\ref{EQN030}) are given in parametric form, where $\rho$ is the 
three-dimensional density of dyons of each kind,
i.e.\ $\rho = K / V_3 = n_D / 2 V_3$ with $V_3$ 
denoting the spatial volume. Hence, $\rho T$ can be interpreted as the 
four-dimensional density of dyons which should actually be viewed as the 
fundamental parameter of the model.

In order to fix physical units, we set the scale by 
choosing $\beta = 1 / T = 1.00 \, \textrm{fm} = 1 / 198 \, \textrm{MeV}$ throughout the paper.
This means, we are considering a certain temperature in the confining phase of SU(2) Yang-Mills theory. 
Setting the string tension to its ``physical value'', taken here as 
$\sigma_\textrm{physical} = (440 \, \textrm{MeV})^2 = 4.99 / \textrm{fm}^2$, 
eqns.\ (\ref{EQN030}) yield a three-dimensional 
density $\rho_0 = 3.10 / \textrm{fm}^3$ 
and $T_c = 389 \, \textrm{MeV}$. 
The dimensionless ratio $T_c / \sqrt{\sigma} = 0.883$ differs from that quoted 
in~\cite{Diakonov:2007nv} by the factor $2^{1/4}$, which originates from the 
replacement $\sigma \rightarrow \sqrt{2} \sigma$ made in~\cite{Diakonov:2007nv} 
in order to take the contribution of antiselfdual dyons into account. 
In general, we define $\bar{d} = (1 / \rho)^{1/3}$, which is proportional to 
and of the same order of magnitude as the average nearest neighbor dyon 
separation in three-dimensional space at density $\rho$. 
In particular, $\bar{d}_0 = (1 / \rho_0)^{1/3} = 0.686 \, \textrm{fm}$ is, 
according to (\ref{EQN030}), the dyon
separation reproducing the physical value of the string tension for 
our choice of temperature, $T = 198 \, \textrm{MeV}$.


\section{\label{sec:SECTION3}Spectral properties of the approximate multi-dyon 
moduli \\ space metric}

We consider selfdual multi-dyon configurations, which are solutions of the 
classical Yang-Mills equations of motion and, therefore, local minima of the 
action functional of Yang-Mills theory. The corresponding surface of minimal 
action can be parameterized by a set of collective coordinates, the dyon positions 
$\mathbf{x}_i^m$ (which have already been introduced in section~\ref{subsec:metric})
and phases which are unimportant for the following. 
This surface is embedded in a flat Euclidean space of infinite dimension spanned 
by the gauge field degrees of freedom $A_\mu^a(x)$. Consequently, the induced moduli 
space metric associated with these collective coordinates \emph{must be positive definite}.

The approximate multi-dyon moduli space metric $g$ proposed in~\cite{Diakonov:2007nv} and the corresponding determinant have been given in  
eqns.\ (\ref{EQN873}) to (\ref{EQN001}).
For the investigation of positive definiteness one has to consider
the eigenvalues of $g$.
In appendix~\ref{SEC011} we show that the number 
of negative eigenvalues of $g$ is equal to four times the number of 
negative eigenvalues of $G$.
Thus it suffices to study $G$:
if $G$ is not positive definite, the same holds for $g$, which means that 
it fails to satisfy the fundamental property of positive definiteness inherent 
to any moduli space metric. In such cases the weight factor $\det(G)$ associated 
with the corresponding dyon configuration cannot be interpreted in a physically 
meaningful way.

Since $G$ is only an approximation of a moduli space metric, 
the existence of loci 
of non-positive definiteness are not excluded. 
Still, one could hope that such cases of failure of the used approximation 
are restricted to a small part of dyon configuration space, which might even 
vanish in the thermodynamic limit. 
Our numerical investigations, however, strongly indicate the opposite, 
i.e.\ that even at small dyon densities the percentage of configurations 
with positive definite $G$ tends to $0$, when the size of the system is 
increased. In the following we present a detailed study of the spectrum 
of $G$, particularly its positive definiteness, for various dyon numbers 
and densities.


\subsection{\label{subsec:analytic_arguments}Analytical considerations}


We start by considering the simple case of a pair of different kind dyons, i.e.\ a caloron. The moduli space metric is exactly 
known~\cite{Kraan:1998pn}. 
At maximally non-trivial holonomy it is given by
\begin{eqnarray}
\label{EQN759} G_d \ \ = \ \ \left(\begin{array}{cc}
2 \pi + 2 / d & -2 / d \\
-2 / d & 2 \pi + 2 / d
\end{array}\right) \, ,
\end{eqnarray}
where $d$ is the dyon separation. It is positive definite with eigenvalues 
\begin{eqnarray}
\label{EQN_lambda_different} \lambda_1 \ \ = \ \ 2 \pi \quad , \quad \lambda_2 \ \ = \ \ 2 \pi + 4 / d \,  .
\end{eqnarray}
Hence, the weight factor $\det(G_d)$ is positive for arbitrary dyon positions. 

Considering, on the other hand, a pair of same kind dyons (using the approximate 
moduli space metric proposed in~\cite{Diakonov:2007nv}) the signs in front of the $2/d$ terms are reversed,
\begin{eqnarray}
\label{EQN013} G_s \ \ = \ \ \left(\begin{array}{cc}
2 \pi - 2 / d & +2 / d \\
+2 / d & 2 \pi - 2 / d
\end{array}\right) \, ,
\end{eqnarray}
yielding eigenvalues
\begin{eqnarray}
\label{EQN_lambda_same} \lambda_1 \ \ = \ \ 2 \pi \quad , \quad \lambda_2 \ \ = \ \ 2 \pi - 4 / d \, .
\end{eqnarray}
Here the weight factor $\det(G_s)$ is positive only for dyon separations 
$d > 2 / \pi \equiv 2 / \pi T$. For our choice of temperature, 
$T = 198 \, \textrm{MeV}$, 
this corresponds to $d > 0.635 \, \textrm{fm}$, which is of the same order of 
magnitude as the average distance of same kind dyons 
$\bar{d}_0 = 0.686 \, \textrm{fm}$ needed to reproduce the physical value of the string tension 
$\sigma_\textrm{physical} = (440 \, \textrm{MeV})^2$ (cf.\ 
section~\ref{subsec:missing_positivity}). 
It has been argued that at the critical $d$ the distance ceases to be a good collective coordinate (see~\cite{Diakonov:2007nv} and references therein) as the dyons strongly overlap.

Of course, such a pair of same kind dyons 
has to be complemented by another pair of the other kind to arrive at the neutral 
situation that we are going to investigate in general
(in other words two-caloron solutions need to be studied, of which a few are known
\cite{Bruckmann:2002vy,Bruckmann:2004nu}). 
Then the moduli space metric becomes 
a $4\times 4$ matrix and the conditions, under which some of its eigenvalues are 
negative, are not that obvious. In section~\ref{subsec:SECT_multi_fixed_rho} 
and~\ref{subsec:SECT_multi_fixed_n} we perform a detailed numerical analysis of 
the dependence of the spectrum of $G$ on the dyon number and density both for 
randomly sampled dyon positions and for dyon positions distributed according 
to $|\det(G)|$. The examples presented here already indicate that the approximate 
moduli space metric $G$ might fail to fulfill the fundamental requirement of 
positive definiteness.


In the following we show that for fixed dyon number $n_D$ the percentage of 
configurations with non-positive definite $G$ becomes larger, when the dyon 
density is increased.
The matrix $G$ can be written in the form
\begin{eqnarray}
\label{EQN_homogen1} G \ \ = \ \ 2 \pi \mathbf{I}_{n_D \times n_D} + D(\mathbf{x}_m^i) \, ,
\end{eqnarray}
where the elements of $D$ are linear combinations of inverse dyon distances 
$1 / d_{i,k}^{m,n}$ (cf.\ eqn.\ (\ref{EQN001})). $D$ is, therefore,  inversely 
homogeneous in the dyon positions, i.e.\
\begin{eqnarray}
\label{EQN_homogen2} D(\mathbf{x}_m^i / \alpha) \ \ = \ \ \alpha D(\mathbf{x}_m^i) \, .
\end{eqnarray}
We start by considering an arbitrary dyon configuration $\{\mathbf{x}_i^m\}$ at 
density $\rho$ fulfilling 
\begin{eqnarray}
\label{inequ} \textrm{Tr}(G) \ < \ 2 \pi n_D \, . 
\end{eqnarray}
For large dyon numbers $n_D$, roughly half of the configurations will 
satisfy this requirement. This is so, 
because $\textrm{Tr}(D)$ is a sum of $(n_D/2)^2$ positive and 
$n_D / 2 \times (n_D / 2 - 1)$ negative terms, i.e.\ it fluctuates approximately 
around zero, and because a traceless $D$ gives exactly the bound of eqn.\ (\ref{inequ}). Obviously, such a matrix $G$ has at least one eigenvalue smaller than $2 \pi$, but might of course be positive definite.

Now we scale all dyon locations $\mathbf{x}^m_i$ by $1 / \alpha < 1$, 
i.e.\ $\{\mathbf{x}_i^m\} \rightarrow \{\mathbf{x}'{}_i^m\} = \{\mathbf{x}_i^m / \alpha\}$, 
which amounts to increasing the dyon density by the factor 
$\alpha^3$, i.e.\ $\rho \rightarrow \rho' = \alpha^3 \rho$. Because of the inverse 
homogeneity of $D$ (eqn.\ (\ref{EQN_homogen2})), the eigenvalues of the squeezed 
configuration, $\lambda'_j$, are related to the eigenvalues of the original 
configuration, 
$\lambda_j$, by multiplying their 
difference to $2 \pi$ by $\alpha$, i.e.\
\begin{eqnarray}
\label{EQN793} \lambda'_j \ \ = \ \ 2 \pi + \alpha (\lambda_j - 2 \pi) \, .
\end{eqnarray}
Consequently, for any eigenvalue $\lambda_j < 2 \pi$ one can choose $\alpha$ 
such that $\lambda'_j < 0$. 
In other words, 
by rescaling any such generic configuration will 
give rise to a configuration for which $G$ is not positive definite.

This argument can be transfered to the 
average spectral density of $G$. The latter is obtained by averaging over randomly 
and uniformly chosen dyon positions inside a cubic spatial volume.
Comparing spectral densities for dyon densities
$\rho$ and $\rho'$ (at fixed dyon number $n_D$) one finds that the latter is 
stretched by the factor 
$\alpha$ around a ``fixed center'' at $2 \pi$ (cf.\ eqn.\ (\ref{EQN793}) and also 
Figure~\ref{FIG005} for numerical evidence, where the center at $2 \pi$ is indicated 
by dashed lines). This illustrates that all eigenvalues smaller than $2 \pi$ will 
eventually become negative, when the dyon density is increased.


\subsection{\label{subsec:SECT_multi_fixed_rho}Multi-dyon configurations at fixed 
density: dependence of the spectrum \\ of $G$ on the dyon number}

In the following we consider dyon configurations with randomly and uniformly chosen 
positions inside a cubic spatial volume $V_3 = L^3$. At fixed density $\rho$ we 
gradually increase the dyon number $n_D$ (and consequently $V_3$), to investigate 
the effect on various quantities characterizing the positive definiteness of $G$:
\begin{itemize}
\item the average percentage $R$ of negative eigenvalues of $G$, 
$R = \langle n_- \rangle / n_D$, 
where $n_-$ denotes the number of negative eigenvalues for a given dyon 
configuration;

\item the probability distribution $P^-(n_-)$ of the number of negative 
eigenvalues of $G$;

\item the spectral density of $G$ (obtained by averaging over sampled dyon 
configurations).
\end{itemize}
Results for two significantly different choices of the dyon density, 
$\rho = 1 / \textrm{fm}^3$ and $\rho = 1 / 125 \, \textrm{fm}^3$, are collected 
in Table~\ref{tab:TABLE1} and Figure~\ref{FIG002}.

\begin{table}[h!]
\centering
\begin{tabular}{|c||c|c|c||c|c|c|}
\hline
 & \multicolumn{3}{|c||}{\vspace{-0.4cm}} & \multicolumn{3}{|c|}{ } \\
 & \multicolumn{3}{|c||}{$\rho = 1 / \textrm{fm}^3$} & \multicolumn{3}{|c|}{$\rho = 1 / 125 \, \textrm{fm}^3$} \\
 & \multicolumn{3}{|c||}{\vspace{-0.4cm}} & \multicolumn{3}{|c|}{ } \\
\hline
 & & & & & & \vspace{-0.4cm} \\
$n_D$ & $L$ in $\textrm{fm}$ & $R$ in $\%$ & $P^-(0)$ in $\%$ & $L$ in $\textrm{fm}$ & $R$ in $\%$ & $P^-(0)$ in $\%$ \\
 & & & & & & \vspace{-0.4cm} \\
\hline
 & & & & & & \vspace{-0.4cm} \\
$100$ & $3.68$ & $21.81(6)$ & $0.000(0)$ & $18.42$ & $0.457(5)$ & $64.53(35)$ \\
$200$ & $4.64$ & $25.04(3)$ & $0.000(0)$ & $23.21$ & $0.546(2)$ & $36.72(2) \ \,$ \\
$400$ & $5.85$ & $28.16(2)$ & $0.000(0)$ & $29.24$ & $0.684(1)$ & $10.22(3) \ \, $ \\
$800$ & $7.37$ & $30.94(1)$ & $0.000(0)$ & $36.84$ & $0.919(4)$ & $\ \, 0.60(40)$\vspace{-0.4cm} \\
 & & & & & & \\
\hline
\end{tabular}
\caption{\label{tab:TABLE1}The average percentage $R$ of negative eigenvalues 
and the 
percentage of dyon configurations with positive definite $G$, $P^-(n_{-}=0)$, 
for various dyon numbers $n_D$ and two selected densities $\rho$ (the averages 
have been computed from $100,000$ independently chosen configurations).}
\end{table}

\begin{figure}[p]
\begin{center}
\input{FIG002.pstex_t}
\end{center}
\caption{\label{FIG002}Histograms obtained from 100,000 independently chosen dyon 
configurations representing the probability distribution $P^-(n_{-})$ of the 
number of negative eigenvalues (as function of $n_{-}/n_D$). 
Left column: $\rho = 1 / \textrm{fm}^3$. 
Right column: $\rho = 1 / 125 \, \textrm{fm}^3$.}
\end{figure}

At first we discuss $\rho = 1 / \textrm{fm}^3$, a density 
smaller by more than a factor of $3$ than the density 
$\rho_0 = 3.10 / \textrm{fm}^3$, which is according 
to~\cite{Diakonov:2007nv} 
needed to produce a quantitatively correct 
string tension, etc. (cf.\ also eqn.\ (\ref{EQN030})). 
The left columns of Table~\ref{tab:TABLE1} 
and Figure~\ref{FIG002} clearly show that for dyon numbers in the range of 
$100 \leq n_D \leq 800$ the probability $P^-(0)$ to find a configurations 
with positive definite $G$, is essentially zero. In other words, among the $100,000$ 
independently chosen dyon configurations there is not a single configuration, 
where the weight factor $\det(G)$ can be interpreted in a physically meaningful 
way, i.e.\ coming from a positive definite moduli space metric. Even worse, 
the average percentage of negative eigenvalues $R$ increases, when we consider 
a larger number of dyons. This implies that a model based on the approximate 
moduli space metric $G$ and 
enforcing its positive definiteness contains highly non-trivial constraints 
determining a small admissible subspace of configuration space, and that the 
restrictiveness of these constraints even increases in the thermodynamic limit.

Since the construction of $G$ in~\cite{Diakonov:2007nv} is based on the 
assumption that all dyons are well separated, one could hope that reducing 
their density might cure the problem.
To check this, we have repeated the analysis for a rather dilute ensemble with 
$\rho = 1 / 125 \, \textrm{fm}^3$. Note that according to the model proposed 
in~\cite{Diakonov:2007nv} this dyon density would yield a rather low value for 
the string tension, 
$\sigma \approx 0.05 \times \sigma_\textrm{physical}$ (cf.\ eqn.\ (\ref{EQN030})). 
Indeed, the average percentage of negative eigenvalues $R$ is now significantly 
smaller, as can be seen in the right 
columns of Table~\ref{tab:TABLE1} and Figure~\ref{FIG002}. However, as before, 
this percentage increases, when we consider a larger number of dyons (keeping
the same density), e.g.\ for $n_D = 800$ dyons less than $1\%$ of all 
configurations have an associated matrix $G$, which is positive definite. 
This gives additional evidence that even for rather low densities the positive 
definiteness of the moduli space metric remains a very selective constraint.

One could still hope that results obtained by an unrestricted integration over 
collective coordinates might evade this problem in the sense that the integrated 
weight associated with dyon configurations with positive definite $G$, 
$W_G^-(n_-=0)$, might be significantly larger than the integrated weights 
corresponding to ``unphysical sectors'', where $G$ is not positive definite, 
namely $W_G^-(n_- \geq 1)$:
\begin{eqnarray}
\label{EQN021} & & \hspace{-0.7cm} W_G^-(n_-) \ \ = \ \ \int \bigg(\prod_{i=1}^K \prod_{m=1}^2 d^3x_i^m\bigg) \, \Delta_{n_-}(\mathbf{x}_i^m) \, \Big|\det(G(\mathbf{x}_i^m))\Big| \, , \\
 & & \hspace{-0.7cm} Z \ \ = \ \ \int \bigg(\prod_{i=1}^K \prod_{m=1}^2 d^3x_i^m\bigg) \, \Big|\det(G(\mathbf{x}_i^m))\Big| \, ,
\end{eqnarray}
where
\begin{eqnarray}
\Delta_{n_-}(\mathbf{x}_i^m) =  \left\{\begin{array}{cl}
1 & \textrm{if }G(\mathbf{x}_i^m)\textrm{ has }n_-\textrm{ negative eigenvalues} \\
0 & \textrm{otherwise}
\end{array}\right. \, .
\end{eqnarray}
We investigate this 
possibility by computing the integrated weights with the absolute value of 
$\det(G)$ and sorting them with respect to the number of negative eigenvalues 
$n_-$. If $W_G^-(0)$ happens to be much larger than $W_G^-(n_-\geq 1)$, 
then the positivity problem might be less severe in actual simulations.
In ref.~\cite{Diakonov:2007nv} this was expected from the strong repulsion 
of dyons at a zero of the metric determinant. 

\begin{figure}[htb]
\begin{center}
\input{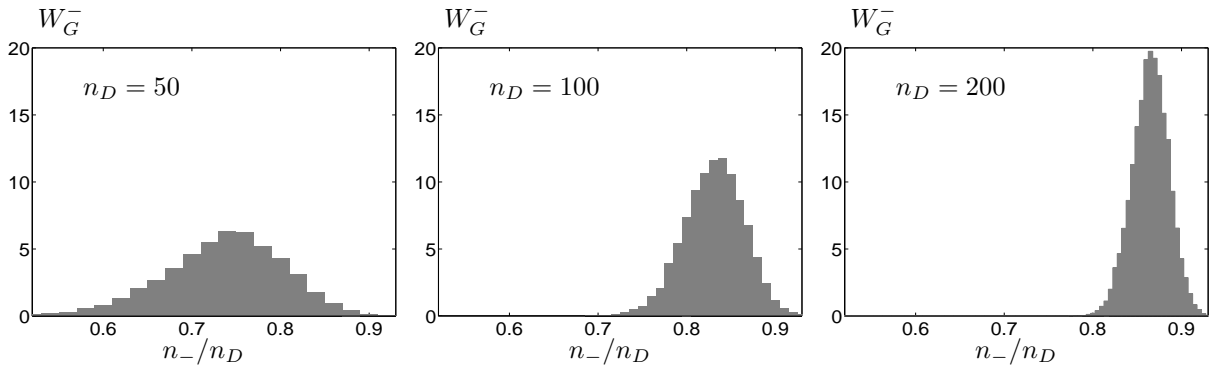}
\end{center}
\caption{\label{FIG003}Integrated weights $W_G^-(n_{-})$ (shown as 
functions of $n_{-} / n_D$) for various dyon numbers $n_D$ and a 
fixed density $\rho = 1 / \textrm{fm}^3$.}
\end{figure}

Since $\det(G)$ exhibits strong fluctuations covering many orders of magnitude, 
we have evaluated (\ref{EQN021}) via Metropolis sampling writing 
$|\det(G)|=\exp(\ln |\det(G)|)$. Results at dyon density 
$\rho = 1 / \textrm{fm}^3$ are shown in Figure~\ref{FIG003}. It is clearly visible, 
that the integrated weight $W_G^-(0)$ of the physically meaningful sector without
negative eigenvalues is negligible compared to the absolute weight of all unphysical
sectors, 
$\sum_{n_- \geq 1} W_G^-(n_-)$. Moreover, one can see that for dyon numbers in 
the range $50 \leq n_D \leq 200$ physical observables are dominated by dyon 
configurations, 
where $G$ has around $80\%$ to $90\%$ negative eigenvalues, and where $W_G^-(n_-)$ 
follows a smooth bell-shaped curve. 

Taking the full determinant into account, i.e.\ taking $\det(G)$ instead 
of $|\det(G)|$, one has to multiply $W_G^-(n_-)$ by an alternating sign 
($+$ if $n_-$ is even, $-$ if $n_-$ is odd). Then roughly half of the dyon 
configurations have associated negative weights, when one computes ensemble 
averages ignoring the requirement of positive definiteness. 
Similar to the case with randomly chosen dyon positions, the 
average percentage $R$ of negative eigenvalues of $G$ increases, when the dyon 
number is increased at fixed density.

We conclude that results based on the approximate moduli space metric $G$ by 
performing an unrestricted integration over collective coordinates (without 
control over positive definiteness) are physically meaningless.


\subsection{\label{subsec:SECT_multi_fixed_n}Multi-dyon configurations at fixed dyon 
number: dependence of the \\ spectrum of $G$ on the density}

To investigate the dependence of the spectrum of $G$ on the dyon density $\rho$, 
we proceed in the same way as in section~\ref{subsec:SECT_multi_fixed_rho}, 
this time keeping the dyon number $n_D$ fixed, while varying the spatial 
volume $V_3$.

\begin{figure}[p]
\begin{center}
\input{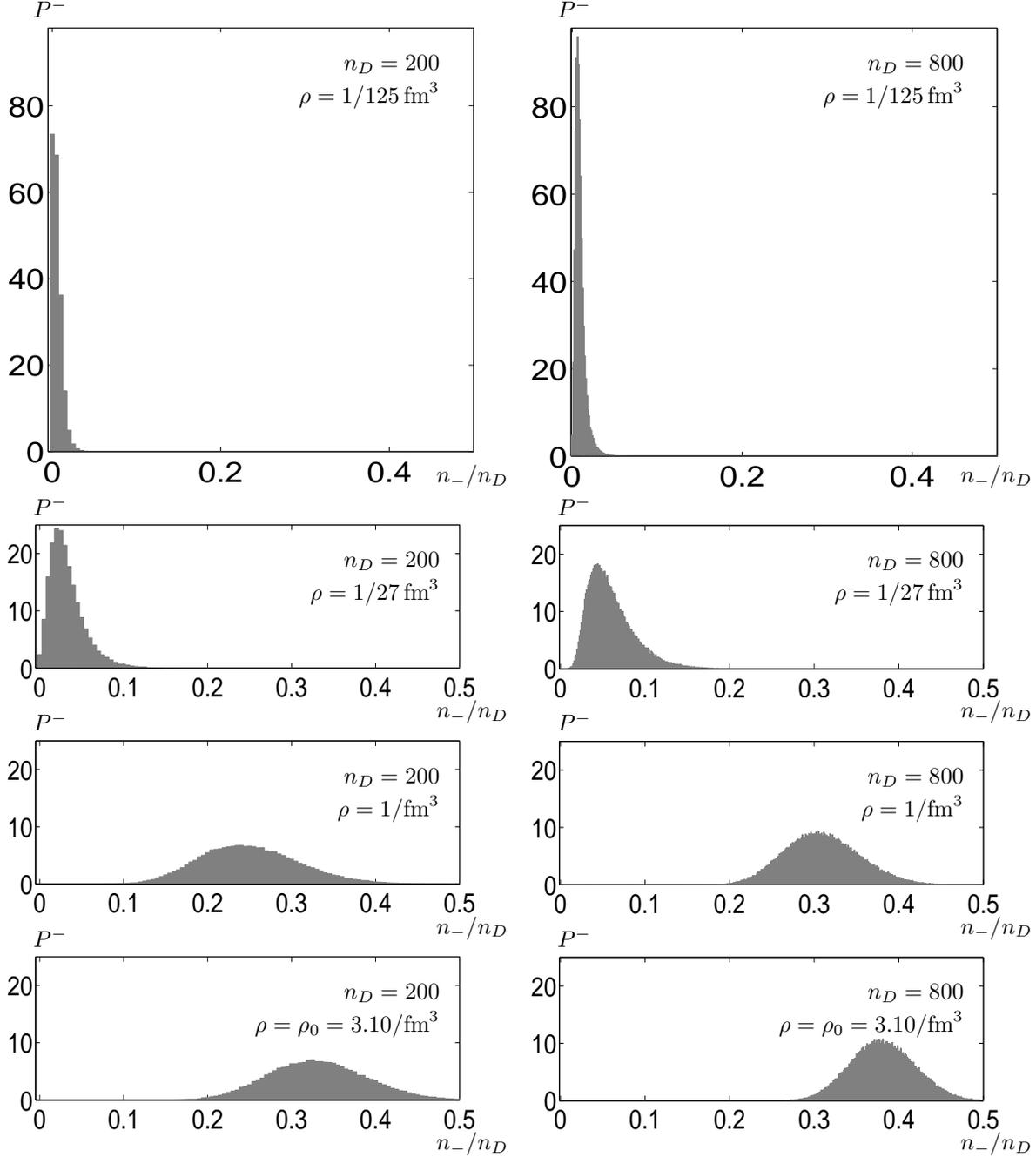}
\end{center}
\caption{\label{FIG004}Histograms obtained from 100,000 independently chosen dyon 
configurations representing the probability distribution $P^-(n_{-})$ of the 
number of negative eigenvalues (as function of $n_{-}/n_D$). 
Left column: $n_D = 200$. Right column: $n_D = 800$.}
\end{figure}

For randomly and uniformly chosen dyon positions, our results show that the average 
percentage of negative eigenvalues $R$ becomes larger, when the density is increased 
(cf.\ Table~\ref{tab:TABLE2} and Figure~\ref{FIG004}). This is in agreement with 
the analytical argument given in section~\ref{subsec:analytic_arguments}. 
This result was expected, since the 
moduli space metric $G$ is an approximation valid only for large dyon separations. 
Even when considering rather dilute ensembles ($\rho = 1 / 125 \, \textrm{fm}^3$), 
still less then $40\%$ of all dyon configurations for $n_D = 200$ 
and less than $1\%$ of all dyon configurations for $n_D = 800$ 
possess a positive definite $G$. 

\begin{table}[h!]
\centering
\begin{tabular}{|c||c|c|c||c|c|c|}
\hline
 & \multicolumn{3}{|c||}{\vspace{-0.4cm}} & \multicolumn{3}{|c|}{ } \\
 & \multicolumn{3}{|c||}{$n_D = 200$} & \multicolumn{3}{|c|}{$n_D = 800$} \\
 & \multicolumn{3}{|c||}{\vspace{-0.4cm}} & \multicolumn{3}{|c|}{ } \\
\hline
 & & & & & & \vspace{-0.4cm} \\
$\rho$ & $L$ in $\textrm{fm}$ & $R$ in $\%$ & $P^-(0)$ in $\%$ & $L$ in $\textrm{fm}$ & $R$ in $\%$ & $P^-(0)$ in $\%$ \\
 & & & & & & \vspace{-0.4cm} \\
\hline
 & & & & & & \vspace{-0.4cm} \\
$1 / 125 \, \textrm{fm}^3$ & $23.21$ & $\ \, \ \, 0.546(2) \ \,$ & $36.72(2) \ \, \ \, \ \,$ & $36.84$ & $\ \, 0.919(4)$ & $0.60(40)$ \\
$1 / 64 \, \textrm{fm}^3$ & $18.57$ & $\ \, \ \, 1.225(8) \ \,$ & $13.28(12) \ \, \ \,$ & $29.47$ & $\ \, 2.305(8)$ & $0.006(4)$ \\
$1 / 27 \, \textrm{fm}^3$ & $13.92$ & $\ \, \ \, 3.241(3) \ \,$ & $1.183(12)$ & $22.10$ & $\ \, 5.997(7)$ & $0.000(0)$ \\
$1 / 8 \, \textrm{fm}^3$ & $\ \, 9.28$ & $\ \, \ \, 9.382(18)$ & $0.003(1) \ \,$ & $14.74$ & $14.60(2) \ \,$ & $0.000(0)$ \\
$1 / \textrm{fm}^3$ & $\ \, 4.64$ & $25.04(3) \ \,$ & $0.000(0) \ \,$ & $\ \, 7.37$ & $30.94(1) \ \,$ & $0.000(0)$ \\
$\rho_0 = 3.10 / \textrm{fm}^3$ & $\ \, 3.18$ & $33.08(2) \ \,$ & $0.000(0) \ \,$ & $\ \, 5.05$ & $38.10(1) \ \,$ & $0.000(0)$\vspace{-0.4cm} \\
 & & & & & & \\
\hline
\end{tabular}
\caption{\label{tab:TABLE2}The average percentage $R$ of negative eigenvalues 
and the percentage of dyon configurations with positive definite $G$, 
$P^-(n_{-}=0)$, for two dyon numbers $n_D$ and various values of the density 
$\rho$ (the averages have been computed from $100,000$ independently chosen 
configurations).}
\end{table}

\begin{figure}[p]
\begin{center}
\input{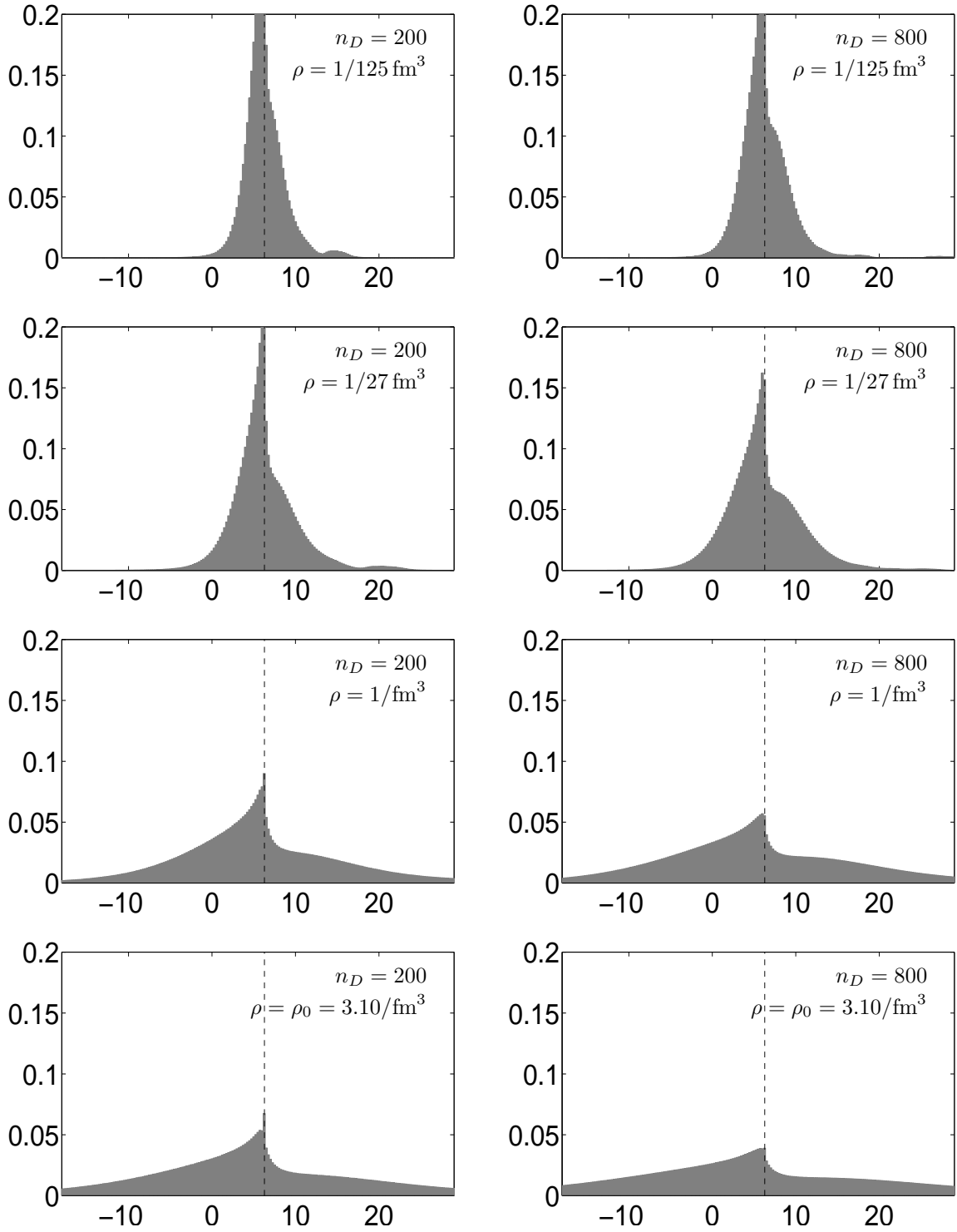}
\end{center}
\caption{\label{FIG005}Histograms obtained from 100,000 independently chosen dyon 
configurations representing the spectral density of the approximate moduli space 
metric $G$. Left column: $n_D = 200$. Right column: $n_D = 800$.}
\end{figure}

To complement the picture, we show in Figure~\ref{FIG005} the spectral density of 
$G$ both for $n_D = 200$ and for $n_D = 800$. From these plots one can clearly 
see that large dyon numbers or densities inevitably yield configurations, where 
the associated matrices $G$ are not positive definite. The scaling of the spectrum 
predicted in section~\ref{subsec:analytic_arguments} is nicely confirmed.

\begin{figure}[htb]
\begin{center}
\input{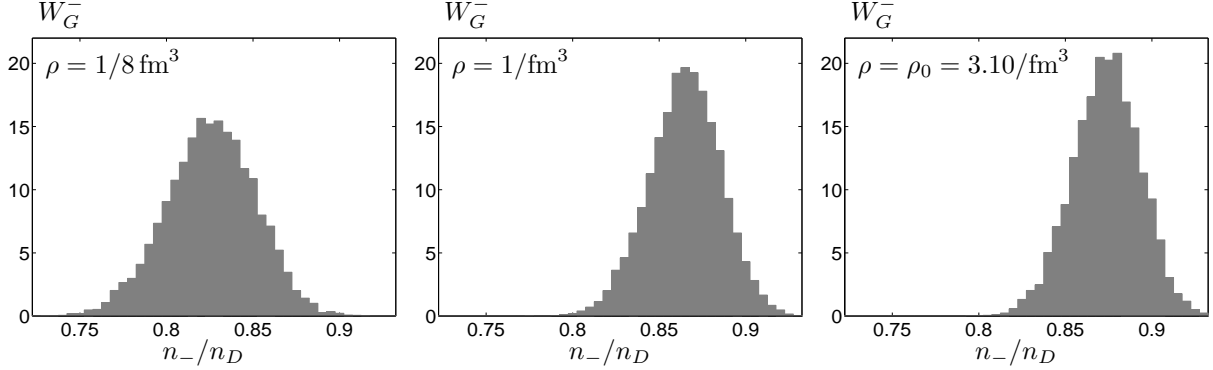}
\end{center}
\caption{\label{FIG006}Integrated weights $W_G^-(n_{-})$, shown as functions of 
$n_{-} / n_D$, as obtained for a number $n_D = 200$ of dyons and various densities.}
\end{figure}

We have also studied the dependence of the integrated weights $W_G^-(n_{-})$ 
(cf.\ eqn.\ (\ref{EQN021})) on the dyon density $\rho$ at fixed dyon number 
$n_D = 200$. Results are shown in Figure~\ref{FIG006}. Again one can see that for 
$\rho \geq 1 / 8 \, \textrm{fm}^3$ observables are dominated by dyon configurations 
with associated matrices $G$, which have between $80\%$ and $90\%$ negative 
eigenvalues. Moreover, the maximum of the integrated weights is shifted towards 
larger numbers of negative eigenvalues, when the dyon density is increased 
towards $\rho_0$ that is considered a realistic value.

\newpage

\section{\label{SEC471}Confinement from a random dyon gas}

In the following we consider dyon ensembles without moduli space metric (or other interactions), i.e.\ we perform a uniform sampling of dyon positions. Although this is a rather extreme simplification of a dyon model (see also~\cite{Bruckmann:2009ne}), such ensembles might be worth studying, because confinement still persists, as we will shortly demonstrate. This investigation might also be helpful to better understand the impact of a moduli space metric on ensembles of dyons.

We consider $n_D = 1,600$ dyons at the same densities as in section~\ref{sec:SECTION3}. We compute the Polyakov loop correlator yielding the free energy of a pair of static charges:
\begin{eqnarray}
\label{EQN779} F_{\bar{Q} Q}(R) \ \ = \ \ -T \ln \Big\langle P(\mathbf{x}) P^\dagger(\mathbf{y}) \Big\rangle \quad , \quad R \ \ = \ \ |\mathbf{x}-\mathbf{y}| .
\end{eqnarray}
The Polyakov loops
\begin{eqnarray}
P(\mathbf{x}) \ \ = \ \ \frac{1}{2} \textrm{Tr}\bigg(\exp\bigg(i \int_0^{1/T} dx_0 \, A_0(\mathbf{x})\bigg)\bigg) \ \ = \ \ \cos\Big(A_0^3(\mathbf{x}) / 2 T\Big) .
\end{eqnarray}
can be evaluated analytically, because the gauge field is treated in the Abelian far field limit (see appendix~\ref{SEC964}). In the algebraic gauge it is given by
\begin{eqnarray}
\label{EQN854} a_0^3(\mathbf{x};q) \ \ = \ \ \frac{q}{r} \quad , \quad a_1^3(\mathbf{x};q) \ \ = \ \ -\frac{q y}{r (r-z)} \quad , \quad a_2^3(\mathbf{x};q) \ \ = \ \ +\frac{q x}{r (r-z)} , 
\end{eqnarray}
where $\mathbf{x} = (x , y , z)$ and $r = |\mathbf{x}|$. All other gauge field components are zero. The multi-dyon configurations we use are linear superpositions of gauge potentials (\ref{EQN854}) determined by dyon positions $\mathbf{x}_i^m$ and charges $q^m$,
\begin{eqnarray}
A_\mu^a(\mathbf{x}) \ \ = \ \ \delta^{a 3} \delta_{\mu 0} \pi T + \sum_{i=1}^K \sum_{m=1}^2 a_\mu^a(\mathbf{x} - \mathbf{x}_i^m;q^m) ,
\end{eqnarray}
where $q^m = \pm 1$ for the $m = 1,2$ dyons plus an additional term in $A_0^3$ generating (in periodic gauge) the non-trivial holonomy $\Omega=\exp(i\pi/2\sigma_3)$ sufficiently far away from all dyons. We regularize the singularities at the dyon centers by factors $1-\exp(-r / \epsilon)$ with
$\epsilon = 0.25 \, \textrm{fm}$ \footnote{We have checked that numerical results remain essentially unaltered, when $\epsilon$ is further decreased.}.

\begin{figure}[htb]
\begin{center}
\input{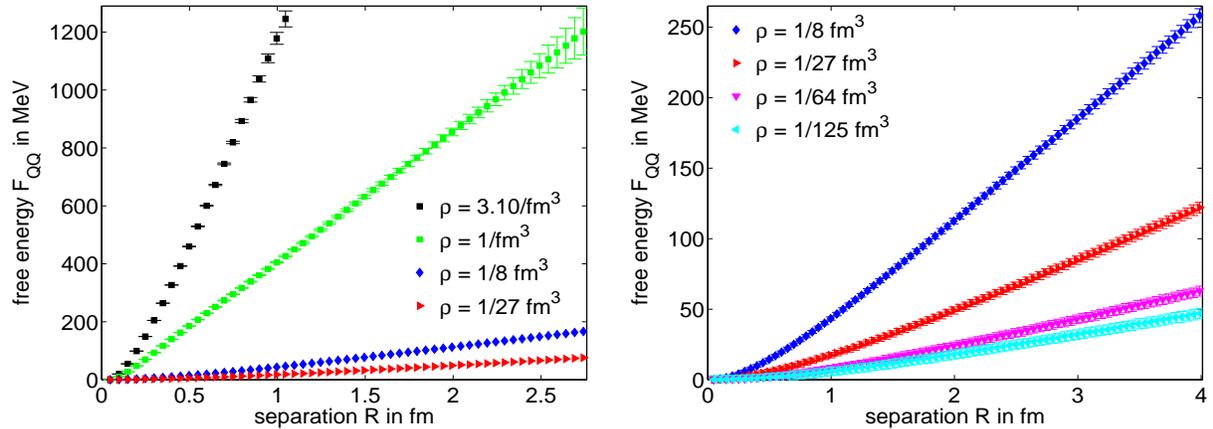}
\end{center}
\caption{\label{FIG007}The free energy of a pair of static charges $F_{Q \bar{Q}}$ as a function of the separation $R$ for various dyon densities $\rho$.}
\end{figure}

As is shown in Figure~\ref{FIG007} the free energy rises linearly up to $\approx 1.5 \, \textrm{fm}$ for $\rho = 3.10 / \textrm{fm}^3$ and up to
$\approx 2.7 \, \textrm{fm}$ for $\rho = 1 / \textrm{fm}^3$,
where statistical noise starts to dominate and up to 
$\approx 4 \, \textrm{fm}$ for $\rho \leq 1 / 8 \, \textrm{fm}^3$, which is the 
maximum separation considered.

The corresponding string tensions are summarized in Table~\ref{tab:TABLE3}.
We conclude that even non-interacting dyons give rise to confinement, when considering the Polyakov loop correlator.
Quantitatively there is a strong dependence of the string tension $\sigma$ on the 
dyon density $\rho$. Note that for dyon density $\rho = \rho_0 = 3.10 / \textrm{fm}^3$ (which is according to the model proposed in~\cite{Diakonov:2007nv} that density yielding approximate agreement with lattice results) the extracted value of the string tension even slightly overshoots its physical value: $\sigma / \sigma^\textrm{physical} \approx 1.4$.

\begin{table}[htb]
\centering
\begin{tabular}{|c|c||c|}
\hline
& & \vspace{-0.4cm} \\
$\rho$ & $L$ in $\textrm{fm}$ & $\sigma / \sigma^\textrm{physical}$ \\
 & & \vspace{-0.4cm} \\
\hline
 & & \vspace{-0.4cm} \\
$1 / 125 \, \textrm{fm}^3$      & $46.42$ & $0.015(1)$  \\
$1 / 64 \, \textrm{fm}^3$       & $37.13$ & $0.020(1)$  \\
$1 / 27 \, \textrm{fm}^3$       & $27.85$ & $0.037(1)$  \\
$1 / 8 \, \textrm{fm}^3$        & $18.57$ & $0.074(1)$  \\
$1 / \textrm{fm}^3$             &  $9.28$ & $0.466(13)$ \\
$\rho_0 = 3.10 / \textrm{fm}^3$ &  $6.37$ & $1.434(19)$ \\
\hline
\end{tabular}
\caption{\label{tab:TABLE3}String tensions in units of the physical string tension $\sigma_\textrm{physical} = (440 \, \textrm{MeV})^2$ for various dyon densities $\rho$.}
\end{table}

On the other hand it is hardly surprising that confinement is already present on such a simple level. Due to the long range nature of the dyon far fields, the models studied in~\cite{Diakonov:2007nv} and also in this paper exhibit certain similarities to ensembles of regular gauge instantons and merons~\cite{Lenz:2003jp,Negele:2004hs,Lenz:2007st} and to the pseudoparticle approach~\cite{Wagner:2005vs,Wagner:2006qn,Wagner:2006du,Szasz:2008qk,Szasz:2008qz}, for which it is well known that confinement and the long range nature of the ``gauge field building blocks'' are closely related (in particular cf.~\cite{Lenz:2007st}, where it has been demonstrated that even a random ensemble of merons yields a linearly rising static potential, as well as~\cite{Wagner:2006qn,Wagner:2006du}, where the relation between the long range nature of the gauge field building blocks and confinement has been established).
In a semiclassical approach at finite temperature, however, dyons are rather the natural building blocks.


\section{\label{sec:SECTION5}Conclusions}

The results obtained in~\cite{Diakonov:2007nv} are based on an approximation
of the multi-dyon moduli space metric, as shown in eqn.\ (\ref{EQN001}).
This metric $G$ -- as we have demonstrated above -- is positive definite only at 
small dyon numbers $n_D$ and small dyon densities $\rho$. 
However, it is used at values of $n_D$ and $\rho$, where it does not satisfy the 
fundamental requirement of positive definiteness. 

In detail our findings are the following:
\begin{itemize}
\item At dyon numbers $100 \leq n_D \leq 800$ and typical densities 
($\rho \approx 1 / \textrm{fm}^3$) practically all dyon configurations correspond 
to matrices $G$, which are not positive definite. 

\item Roughly half of the dyon configurations have odd numbers of negative 
eigenvalues. This implies that every second dyon configuration receives a negative
weight factor, when the metric determinant is taken into account as weight of an 
unrestricted integration over collective coordinates.

\item All attempts to approach the thermodynamic limit by increasing the dyon 
number, while keeping their density fixed, have lead to more severe violations 
of positive definiteness.

\item Decreasing the temperature (throughout this work we have used 
$\beta = 1 / T = 1.00 \, \textrm{fm}$), 
while keeping the physical values of the string tension and the critical 
temperature fixed (according to eqn.\ (\ref{EQN030})), amounts to increasing 
the three-dimensional density. As we have demonstrated this makes the 
situation even worse.

\item Dyons generically induce confinement; already dyon ensembles with randomly chosen positions do so.

\end{itemize}

We expect that the problems encountered for SU(2) are generic also for higher gauge groups.

We consider these findings a challenge.
Since positive definiteness is a fundamental property of any moduli space metric, 
it seems doubtful that results without restriction to the ``positive definite 
subset'' of dyon configurations can be interpreted in a physically meaningful 
way. If the metric determinant is taken into account as weight,
averages of physical observables are computed from alternating sums over
the number of negative eigenvalues of $G$.
We conclude that, in order to obtain physically meaningful results, it is 
necessary to either modify the dyon model in a way that the (corrected) weight 
factor is always positive, or to restrict the integration over dyon positions 
to those parts of configuration space, where $G$ is positive definite.

In a subsequent publication we plan to 
present numerical simulations of dyon models with interactions derived from the approximate moduli space metric $G$, which do not suffer from ``unphysical sign problems''. One appealing possibility is the 
following integration over collective coordinates:
\begin{eqnarray}
\bigg(\prod_{i=1}^K \prod_{m=1}^2 d^3x_i^m\bigg) W \quad , \quad W \ \ = \ \ \left\{\begin{array}{cl} \det(G) & \textrm{if }G\textrm{ is positive definite} \\ 0 & \textrm{otherwise}\end{array}\right. \, .
\end{eqnarray}
Since the constraints imposed on the dyon coordinates $\mathbf{x}_i^m$ by this measure are highly non-trivial, it seems unlikely that this model can be treated analytically. Therefore, we are currently developing efficient Monte-Carlo algorithms, which are respecting the positive definiteness of $G$, i.e.\ which are designed to avoid ``forbidden'' multi-dyon configurations. In this respect our finding that already dyons with randomly chosen positions generate confinement is 
encouraging in the sense that models based on dyons seem to capture
the relevant degrees of freedom of SU(2) Yang-Mills theory.


\appendix


\section{\label{SEC011}Relation between the spectrum of $G$ and the spectrum of $g$}

In the following we show that the number $n_-(g)$ of negative eigenvalues of $g$ 
is four times the number $n_-(G)$ of negative eigenvalues of $G$. 

The relation between $G$ and $g$ is
\begin{eqnarray}
\label{EQN630}  
\nonumber & & \hspace{-0.7cm} g \ \ = \ \ \left(\begin{array}{cccc}
G + W_x^T G^{-1} W_x & W_x^T G^{-1} W_y & W_x^T G^{-1} W_z & W_x^T G^{-1} \\
W_y^T G^{-1} W_x & G + W_y^T G^{-1} W_y & W_y^T G^{-1} W_z & W_y^T G^{-1} \\
W_z^T G^{-1} W_x & W_z^T G^{-1} W_y & G + W_z^T G^{-1} W_z & W_z^T G^{-1} \\
G^{-1} W_x & G^{-1} W_y & G^{-1} W_z & G^{-1}
\end{array}\right) \ \ = \\
\label{EQN631} & & = \ \ 
\underbrace{\left(\begin{array}{cccc}
1 & 0 & 0 & W_x^T \\
0 & 1 & 0 & W_y^T \\
0 & 0 & 1 & W_z^T \\
0 & 0 & 0 & 1
\end{array}\right)}_{= \tilde{W}^T}
\underbrace{\left(\begin{array}{cccc}
G & 0 & 0 & 0 \\
0 & G & 0 & 0 \\
0 & 0 & G & 0 \\
0 & 0 & 0 & G^{-1}
\end{array}\right)}_{= \tilde{G}}
\underbrace{\left(\begin{array}{cccc}
1 & 0 & 0 & 0 \\
0 & 1 & 0 & 0 \\
0 & 0 & 1 & 0 \\
W_x & W_y & W_z & 1
\end{array}\right)}_{= \tilde{W}}
\end{eqnarray}
cf.\ eqn.\ (20) in~\cite{Diakonov:2007nv}.

Obviously, $\tilde{G}$ has $4 n_-(G)$ negative eigenvalues. We denote the eigenvectors 
of $\tilde{G}$ by $\mathbf{v}^{(k,\pm)}$ and the eigenvalues by 
$\lambda^{(k,\pm)}$, i.e.\ $\tilde{G} \mathbf{v}^{(k,\pm)} = \lambda^{(k)} \mathbf{v}^{(k,\pm)}$. 
$+$ denotes a positive eigenvalue, i.e.\ $\lambda^{(k,+)} > 0$, and $-$ denotes a negative 
eigenvalue, i.e.\ $\lambda^{(k,-)} < 0$. Because $\tilde{G}$ is symmetric, the eigenvectors 
can be chosen orthonormal. Moreover, all eigenvalues are real and all eigenvectors can be 
chosen real.

With arbitrary real coefficients $\alpha(k)$ one has
\begin{eqnarray}
\nonumber & & \hspace{-0.7cm} 0 \ \ > \ \ \sum_k \alpha(k)^2 \lambda^{(k,-)} \ \ = \ \ \bigg(\sum_k \alpha(k) \mathbf{v}^{(k,-)}\bigg)^T \tilde{G} \bigg(\sum_{k'} \alpha(k') \mathbf{v}^{(k',-)}\bigg) \ \ = \\
\label{EQN797} & & = \ \ \bigg(\tilde{W}^{-1} \sum_k \alpha(k) \mathbf{v}^{(k,-)}\bigg)^T \underbrace{\tilde{W}^T \tilde{G} \tilde{W}}_{= g} \bigg(\tilde{W}^{-1} \sum_{k'} \alpha(k') \mathbf{v}^{(k',-)}\bigg) .
\end{eqnarray}
%
The $4 n_-(G)$ dimensional subspace
\begin{eqnarray}
S \ \ = \ \ \tilde{W}^{-1} \sum_k \alpha(k) \mathbf{v}^{(k,-)}
\end{eqnarray}
can be expanded in terms of the orthonormal and real eigenvectors $\mathbf{x}^{(k,\pm)}$ 
of $g$ ($g$ is also symmetric), i.e.\
\begin{eqnarray}
S \ \ = \ \ \sum_k \beta(k,-) \mathbf{x}^{(k,-)} + \sum_k \beta(k,+) \mathbf{x}^{(k,+)} .
\end{eqnarray}
If $g$ would have less than $4 n_-(G)$ negative eigenvalues, one could chose a combination 
of $\alpha(k)$ such that $\beta(k,-) = 0$. This, however, would be a contradiction 
to (\ref{EQN797}). Therefore, $g$ has at least $4 n_-(G)$ negative eigenvalues, 
i.e.\ $n_-(g) \geq 4 n_-(G)$.

Analogously one can show that $n_+(g) \geq 4 n_+(G)$. This proves that $n_-(g) = 4 n_-(G)$.


\section{\label{SEC964}The dyon gauge field in the far field limit}

The gauge field of a single dyon can be obtained by considering the gauge field of a caloron at maximal holonomy in the limit of infinite dyon separation.

When the distance to the dyon center $r = (x^2 + y^2 + z^2)^{1/2}$ is large, 
the gauge field $A_\mu = A_\mu^a \sigma^a / 2$ is Abelian:
\begin{eqnarray}
A_\mu^1 \ \ = \ \ 0 \quad , \quad A_\mu^2 \ \ = \ \ 0 \quad , \quad A_\mu^3 \ \ = \ \  \delta_{\mu 0} \pi T + q \bar{\eta}_{\mu \nu}^3 \partial_\nu \ln(\phi) ,
\end{eqnarray}
where
\begin{eqnarray}
\phi \ \ = \ \ \frac{1}{r - z}
\end{eqnarray}
and $\bar{\eta}_{\mu \nu}^a = \epsilon_{a \mu \nu} - \delta_{a \mu} \delta_{0 \nu} + \delta_{a \nu} \delta_{0 \mu}$ is the t'Hooft symbol. 
The charge $q$ is either $+1$ (dyons of the first kind) or $-1$ (dyons of the second kind). 
The coordinate system has been chosen such, that the singular Dirac string points in 
positive $z$-direction.

The non-zero components of the gauge field read
\begin{eqnarray}
\label{EQN679} A_0^3 \ \ = \ \ \pi T + \frac{q}{r} \quad , \quad A_1^3 \ \ = \ \ -\frac{q y}{r (r-z)} \quad , \quad A_2^3 \ \ = \ \ +\frac{q x}{r (r-z)} .
\end{eqnarray}
With the definitions $E_j = F_{0 j}$, $B_j = -\epsilon_{j k l} F_{k l} / 2$
and $F_{\mu \nu} = \partial_\mu A_\nu^3 - \partial_\nu A_\mu^3$ the corresponding 
electric and magnetic monopole fields are given by
\begin{eqnarray}
\mathbf{E} \ \ = \ \ \frac{q \mathbf{r}}{r^3} \quad , \quad \mathbf{B} \ \ = \ \ \frac{q \mathbf{r}}{r^3} .
\end{eqnarray}

Throughout this paper we approximate dyons always by these Abelian far fields.

\section*{Acknowledgments}

We thank Dmitri Diakonov, Hilmar Forkel and Carsten Urbach for helpful discussions.
F.B.\ acknowledges support by DFG (BR 2872/4-1). This work has been supported 
in part by the DFG Sonderforschungsbereich/Transregio SFB/TR9-03.



\end{document}